\RequirePackage{ifpdf}
\ifpdf 
\documentclass[pdftex]{sigma}
\else
\documentclass{sigma}
\fi

\begin{document}
\allowdisplaybreaks

\renewcommand{\PaperNumber}{038}

\FirstPageHeading

\renewcommand{\thefootnote}{$\star$}

\ShortArticleName{Nonlinear Fokker--Planck  Equation in the Model
of Asset Returns}

\ArticleName{Nonlinear Fokker--Planck  Equation\\ in the Model of
Asset Returns}

\Author{Alexander SHAPOVALOV~$^{\dag\ddag\S}$, Andrey
TRIFONOV~$^{\ddag\S}$ and Elena MASALOVA~$^{\ddag}$}
\AuthorNameForHeading{A. Shapovalov, A. Trifonov and E. Masalova}

\Address{$^\dag$~Tomsk State University, 36 Lenin Ave., 634050
Tomsk, Russia}
\EmailD{\href{mailto:shpv@phys.tsu.ru}{shpv@phys.tsu.ru}}

\Address{$^\ddag$~Tomsk Polytechnic University, 30 Lenin Ave.,
634050 Tomsk, Russia}
\EmailD{\href{mailto:trifonov@mph.phtd.tpu.edu.ru}{trifonov@mph.phtd.tpu.edu.ru}, \href{mailto:eash@mail2000.ru}{eash@mail2000.ru}}

\Address{$^\S$~Mathematical Physics Laboratory, Tomsk
Polytechnic University,\\
$\phantom{^\S}$~30 Lenin Ave., 634050 Tomsk, Russia}

\ArticleDates{Received September 30, 2007, in f\/inal form March
26, 2008; Published online April 06, 2008}

\Abstract{The Fokker--Planck equation with dif\/fusion
coef\/f\/icient quadratic in space va\-riab\-le, linear drift
coef\/f\/icient,
 and nonlocal nonlinearity term is
considered  in the framework of a model of analysis of asset
returns at f\/inancial markets. For special cases of such
a~Fokker--Planck equation we describe a construction of exact
solution of the Cauchy problem. In the general case, we construct
the leading term of the Cauchy problem solution asymptotic in
a~formal small parameter in semiclassical approximation following
the complex WKB--Maslov method in the class of trajectory
concentrated functions.}


\Classification{35K55; 62M10; 91B28; 91B84}

\section{Introduction}
\label{Introduc}

Studies of observations available at f\/inancial markets reveal
empirically  stated statistical regularities of asset price
changes whose distributions are of broad interest providing more
reasons to control of losses and gains in the course of
quotations. Asset returns  can be measured in terms of increments
$\Delta\xi=\xi(t+\Delta t)-\xi(t)$ over the time interval $\Delta
t$ of a  time series  $\xi(t)$ of quotations of an asset. The
quantitative handling of high-frequency data of currency exchange
rates  (see \cite{FRIEDRICH} and references therein) have
substantiated  heavy tailed probability density functions of price
increments $\Delta \xi$ for small time steps $\Delta t$,
asymptotic power law (or heavy tails) of return distributions
dif\/ferent from the Gaussian distribution,  volatility
clustering, etc.

Power law is considered to be well-established fact (see, e.g.,
\cite{MANTEGNA}). Several semi-empirical mathematical models have
been proposed to describe the non-Gaussian properties of asset
return distributions. The most known  of them are  ARCH (auto
regression conditional  heteroskedastic models with time changing
conditional variances) \cite{ENGLE} and its modif\/ication GARCH
\cite{BOLLERSEV}, ``truncated'' Levy  distributions
\cite{MANTEG-STAN}, multiplicative noise models (see, e.g.,
\cite{VRIES}), multifractal cascade models with the so called
HARCH ef\/fect  \cite{MUZY}.

The model proposed in \cite{FRIEDRICH} to describe asset return
distributions uses the approach based on the Fokker--Planck
equation (FPE) with variable coef\/f\/icients. In some papers this
equation is called the Fokker--Planck--Kolmogorov equation. The
idea of the work \cite{FRIEDRICH} is based on similarity
observation between asset return distributions and distributions
of vortex velocity increments in turbulent cascades in
hydrodynamics. This observation is verif\/ied by quantitative
analysis of distribution parameters that justif\/ies an analogy
between turbulent cascade in a stream of liquid and information
cascade at stock markets.

Following this semi-empirical analogy, we f\/ix our  attention on
the power law of the probability density function (PDF) on one
hand, and on the analogy with the turbulent cascade, on the other.
For such a system,  self-organization phenomena are the
characteristics that could be mathematically expressed by an
additional nonlinear mean drift term in the FPE \cite{FRANK}.

In this work we modify the FPE with the dif\/fusion
coef\/f\/icient quadratic in $\Delta\xi$  considered in
\cite{FRIEDRICH} by introducing the nonlinear mean drift term. For
such a FPE, we apply the approach developed in  \cite{SHIINO,
BEL-TRIFONOV, SH-REZ-TR} which reduces the integrability problem
for the nonlinear FPE to the problem for a linear equation
associated  with the original nonlinear FPE. We construct the
exact solution of the nonlinear FPE whose coef\/f\/icients have  a
special form, and for the coef\/f\/icients of a~general form we
construct the leading term of solution asymptotic in a~formal
small parameter in semiclassical approximation in the class of
trajectory concentrated functions.

\section[Formalism of the Fokker-Planck equation]{Formalism of the Fokker--Planck equation}

\label{NLFPK-aa}

The formalism, proposed in \cite{FRIEDRICH} to describe
distributions  of stock market returns which asymp\-to\-tical\-ly
obey  power law, is  based on the Fokker--Planck equation in the
evolution variable $\Delta t$. The power law exponents are
determined from the Kramers--Moyal coef\/f\/icients founded in the
framework of a phenomenological method developed in
\cite{FRIEDRICH}. The FPE provides the information as to how the
distributions on dif\/ferent time intervals $\Delta t$ are
correlated. In view of the analogy with the turbulent cascade the
FPE is written in \cite{FRIEDRICH} in terms of logarithmic
time-scale cascade proceeds.

The FPE for the probability density function $u(\Delta\xi,\tau)$
of price increment $\Delta\xi$ at a given logarithmic time scale
$\tau$ is derived in \cite{FRIEDRICH} in the standard way from the
Chapman--Kolmogorov equation and is presented in the form
\begin{gather}
 \partial_\tau u(\Delta\xi,\tau )=\left[- \frac{\partial}{\partial(\Delta \xi)}D^{(1)}(\Delta\xi,\tau)
+
\frac{\partial^2}{\partial(\Delta\xi)^2}D^{(2)}(\Delta\xi,\tau)\right]u(\Delta\xi,\tau
). \label{FPK-FRIED}
\end{gather}
Here $\partial_\tau=\partial/\partial\tau$, and the Kramers--Moyal
coef\/f\/icients, the drift $D^{(1)}(\Delta\xi,\tau)$ and
dif\/fusion $D^{(2)}(\Delta\xi,\tau)$ coef\/f\/icients, are found
directly from the f\/inancial data and the following
approxi\-mations were obtained
\begin{gather}
 D^{(1)}(\Delta\xi ,\tau)=-0.44 \Delta\xi, \label{D1} \\
 D^{(2)}(\Delta\xi ,\tau)=0.003 e^{-\tau/2} + 0.019(\Delta\xi +0.04)^2. 
\label{D-k}
\end{gather}

We assume the solutions $u(\Delta\xi,\tau)$ of equation
(\ref{FPK-FRIED}) to be real smooth functions decreasing as
$|\Delta\xi|\to\infty$. Notice, that equation (\ref{FPK-FRIED})
can be written as a balance relation that makes obvious the
conservation of the integral $\int
_{-\infty}^{+\infty}u(\Delta\xi,\tau) d(\Delta\xi)$. Taking the
initial function   $u(\Delta \xi,0)$ to be normalized, we have
\begin{gather}
\label{norm} \int_{-\infty}^{+\infty}u(\Delta\xi,\tau)
d(\Delta\xi)=1, \qquad \tau\geqslant 0.
\end{gather}
The linear FPE (\ref{FPK-FRIED}), (\ref{D1}), (\ref{D-k}) is
solved numerically in \cite{FRIEDRICH} and quantitative agreement
is obtained between the solutions and distributions constructed
directly from the data of quotations, i.e.\ experimental data.

A more detailed analysis of the  power laws $u(\Delta\xi)\thicksim
(\Delta\xi)^{-(1+\mu)}$
made in \cite{SORNETTE} shows that the exponents $\mu$ found in
the framework of the FPE formalism are not in good agreement with
experimentally obtained values.

Dif\/ferent ideas have been involved in studying of ``heavy
tails'' of f\/inancial time-series distributions. Mathematically,
such  nonlinear stochastic systems are described in terms of the
Fokker--Planck equation with coef\/f\/icients dependent on the
distribution. We consider the feedback caused by reciprocal
ef\/fects of assets at the f\/inancial market. In accordance with
\cite{FRANK} and by analogy between the exchange rates and the
hydrodynamic turbulence \cite{FRIEDRICH, SORNETTE} let us modify
the FPE (\ref{FPK-FRIED}) assuming that its coef\/f\/icients
$D^{(1)}$ and $D^{(2)}$ depend on the order parameter  described
by the f\/irst moment
\begin{gather}
\label{moment}  X_u(\tau)= \int_{-\infty}^{\infty}\Delta \xi
u(\Delta \xi,\tau)d(\Delta \xi).
\end{gather}

In this work we consider analytical methods of solution
construction for the following  special case of the nonlinear
Fokker--Planck equation (NFPE)
\begin{gather}
  \partial_\tau u(\Delta\xi,\tau )=\left[ \frac{\partial}{\partial(\Delta\xi)}\left(\alpha\Delta \xi+\varkappa  X_u(\tau)\right)
+ \frac{\partial}{\partial(\Delta\xi)}D^{(2)}(\Delta\xi,\tau)
\frac{\partial}{\partial(\Delta\xi)}\right]u(\Delta\xi,\tau ).
\label{FPK-NL-2}
\end{gather}
Here $\alpha$ and  $\varkappa$ are real positive parameters. The
dif\/fusion coef\/f\/icient  $ D^{(2)}$  is taken in more general
form  as compared to (\ref{D-k}), i.e.
\begin{gather}
D^{(2)}(\Delta\xi, \tau)= \epsilon\big( f(\tau)+ \big(a \Delta\xi
+ g(\tau)\big)^2  \big). \label{D-2}
\end{gather}
Here $f(\tau)(\geqslant 0)$ and  $g(\tau)$ are real smooth
functions, $a$ is a real parameter, $\epsilon$ is a real positive
parameter.

Market impact on price increment  $\Delta \xi$ of an asset with
the probability density function $u(\Delta\xi,\tau)$ determined by
(\ref{FPK-NL-2}), is modeled by the coef\/f\/icient $\varkappa
X_{u}(\tau)$.

The motivation of the drift coef\/f\/icient
\begin{gather}
D^{(1)}(\Delta\xi, \tau)= -\alpha \Delta\xi -\varkappa X_u(\tau)
\label{D-1}
\end{gather}
in (\ref{FPK-NL-2}) can be  taken from the stochastic methods
\cite{GARDINER} and empirical data handling \cite{FRIEDRICH,
SORNETTE}. In the Fokker--Planck equation
\begin{gather}
\label{FP-V}
\partial_\tau u(\Delta\xi,\tau)=\partial_{\Delta\xi}\big[ -V_{\Delta\xi}(\Delta\xi,\tau)+D^{(2)} \partial_{\Delta\xi}   \big]u(\Delta\xi,\tau)
\end{gather}
the function $V(\Delta\xi,\tau)$  can be treated as a  potential
of a regular external force
$V_{\Delta\xi}(\Delta\xi,\tau)=\partial_{\Delta\xi}
V(\Delta\xi,\tau)$ \cite{GARDINER}. Let us assume that there are a
number of mutually interacting  assets  on  the f\/inancial
market. Following the mean f\/ield concept \cite{FRANK}, we modify
the FPE  (\ref{FP-V})  by  substitution
\begin{gather}
V_{\Delta\xi}(\Delta\xi,\tau)\rightarrow
V_{\Delta\xi}(\Delta\xi,\tau) +
\int_{-\infty}^{\infty}W_{\Delta\xi}(\Delta\xi-y,\tau) u(y,\tau)
dy. \label{POTENTIAL}
\end{gather}
Here the integral simulates the cumulative ef\/fect of other
market assets on price increment $\Delta\xi$ of a separate asset
with a potential  $W(\Delta\xi-y,\tau)$ which depends on the
dif\/ference $\Delta\xi-y$  due to symmetry of mutual asset
ef\/fects.

In accordance with \cite{FRIEDRICH,SORNETTE}, where linear   form
of the drift  $V_{\Delta\xi}(\Delta\xi,\tau)=\tilde\alpha
\Delta\xi$ gives close approximations of empirical data, we  take
modif\/ied drift coef\/f\/icient (\ref{POTENTIAL}) as
$-\tilde\alpha \Delta\xi+\varkappa \int_{-\infty}^{\infty}
(\Delta\xi-y)u(y,\tau) d y$ that results in $-\alpha \Delta\xi
-\varkappa X_u(\tau)$, $\alpha=\tilde\alpha -\varkappa$, that, in
turn, gives (\ref{D-1}).

To estimate the parameter $\varkappa$,   one can  take into
account that $\varkappa X_u(\tau)$ in  (\ref{D-1}) is responsible
for cumulative ef\/fect of all  marketable assets  on the price
increment $\Delta \xi$ of a given asset. It is assumed that in the
stock market $\varkappa X_u(\tau)$ can be estimated with the help
of an appropriate market index,  as it characterizes the market as
a whole. In the  exchange  operations (FOREX market), where
indices are not used, such a characteristics could be deduced from
the correlation matrix of currency exchange rates. Anyway, the
estimation of the parameter $\varkappa$ is a subject of a~special
study.

The integrability problem for the FPE  in applications is
practically solved numerically. As a rule, analytical approaches
are based on special classes of functions in which solutions are
constructed. A more detailed review of solution  methods for the
NFPE (\ref{FPK-NL-2}) can be found in \cite{SH-REZ-TR}.

Using the approach considered in \cite{SH-REZ-TR} we obtain the
evolution operator for equation (\ref{FPK-NL-2}) with a constant
dif\/fusion coef\/f\/icient $D^{(2)}={\mbox{\rm const}}$. This
result is  used to construct exact solutions of  (\ref{FPK-NL-2})
for the functions $f(\tau)$ and $g(\tau)$ of a special form.
Following \cite{BEL-TRIFONOV} for these functions of general form,
we construct asymptotic solutions of (\ref{FPK-NL-2}) in
semiclassical approximation in a small parameter $\epsilon\to 0$
in the class of trajectory concentrated functions.

\section[Evolution operator for the FPE with constant diffusion]{Evolution operator for the FPE with constant dif\/fusion}
\label{NLFPK}

Let us assume that in the NFPE (\ref{FPK-NL-2}) the dif\/fusion is
constant,  $D^{(2)}(\Delta x,\tau)=\epsilon$.  Setting
$\Delta\xi=x$ for convenience, we  rewrite  (\ref{FPK-NL-2}) as
\begin{gather}
 \partial_\tau u( x,\tau )=\big[ \partial_x\left(\alpha x+\varkappa X_u(\tau)\right) +\epsilon\partial_{xx} \big]u(x,\tau ),
\label{FPK-NL-3}
\end{gather}
where $\partial_x=\partial/\partial x$,
$\partial_{xx}=\partial^2/(\partial x)^2$. Equation
(\ref{FPK-NL-3}) plays the  basic role in our study of the
nonlinear Fokker--Planck equation (\ref{FPK-NL-2}) and in  the
linear case ($\varkappa=0$) it describes the Ornstein--Uhlenbeck
process \cite{UHLENBECK} (see also~\cite{GARDINER}).

Let us take a f\/ixed value $\tau=s$ as the initial time and
consider the Cauchy problem
\begin{gather}
\label{CAUCHY-1} u( x,s)=\gamma( x)
\end{gather}
for equation (\ref{FPK-NL-3}) in a class of functions of the
Schwartz space $\mathcal{S} $ in the variable  $x$ for
$\tau\geqslant s$. In accordance with (\ref{norm})  we can set
\begin{gather*} 
\gamma( x)\in \mathcal{S}, \qquad \int_{-\infty}^{+\infty} \gamma(
x)d x=1
\end{gather*}
without the loss of generality then
\begin{gather}
\label{NORM-1} \int_{-\infty}^{+\infty}u( x,\tau)d  x=1, \qquad
\tau\geqslant s.
\end{gather}
For the f\/irst moment (\ref{moment}) of the function $u(
x,\tau)$,
\begin{gather}
\label{drift} X_u(\tau)= \int_{-\infty}^{+\infty}x u( x,\tau)d x,
\end{gather}
from   (\ref{FPK-NL-3}) and  (\ref{NORM-1}) we immediately obtain
\begin{gather}
\label{EE-1} \dot{X}_u(\tau)=-(\alpha+\varkappa ) X_u(\tau).
\end{gather}
Here $\dot{ X}_u(\tau)=d X_u(\tau)/d \tau$, then
\begin{gather}
\label{EE-2} X_u(\tau)=X_\gamma\exp\left[-(\alpha+\varkappa
)(\tau-s)\right],
\\
\label{EE-3} X_\gamma= \int_{-\infty}^{+\infty}x \gamma( x)d x .
\end{gather}
Therefore, the f\/irst moment $X_u(\tau)$ is determined by the
initial distribution $\gamma(x)$ and does not require the solution
of equation (\ref{FPK-NL-3}). The substitution of (\ref{EE-2})
into (\ref{FPK-NL-3}) actually linearizes the NFPE. Such a
property of the NFPE with coef\/f\/icients depending on the
f\/irst moment $X_u(\tau)$ is related to the fact that this
equation is close to the linear one (see \cite{SH-REZ-TR} for more
details).

Let us present the solution of the Cauchy problem (\ref{FPK-NL-3})
and (\ref{CAUCHY-1}) as action of the evolution operator $\widehat
U (\tau, s, \cdot)$ on the initial function  $\gamma (x)$.
Following \cite{SH-REZ-TR} we obtain
\begin{gather}\label{GREEN-NLIN}
u(x,\tau)=\widehat U (\tau, s, \gamma)(x)=
\int_{-\infty}^{+\infty}G_{\rm lin} (\tau,s, x- X_u(\tau), y-
X_\gamma) \gamma( y)d y.
\end{gather}
Here  $G_{\rm lin} (\tau,s, x, y)$ is the Green function of the
linear FPE  resulting from (\ref{FPK-NL-3}) with the assumption
$\varkappa=0$,
\begin{gather*}
G_{\rm lin} (\tau,s, x, y)=
\frac{e^{\frac{\alpha}{2}(\tau-s)}}{\sqrt{\frac{4\pi\epsilon}{\alpha}\sinh\alpha(\tau-s)}}
\exp\left[-
 \frac{\alpha e^{\alpha(\tau-s)}}{4\epsilon\sinh\alpha(\tau-s)}\big( x-e^{-\alpha(\tau-s)}y \big)^2  \right].
\end{gather*}

Notice that the solution of the Cauchy problem  (\ref{FPK-NL-3})
and  (\ref{CAUCHY-1}) can be immediately generalized for the
multidimensional FPE.

\section[Nonlinear FPE with quadratic diffusion]{Nonlinear FPE with quadratic dif\/fusion}
\label{NLFPK-1}

Consider the NFPE
\begin{gather}
  \partial_\tau u( x,\tau )=\big[\partial_x \left(\alpha x+\varkappa \beta(\tau)\right) +\epsilon\partial_{x} \big(f(\tau)+\big(a x+
g(\tau)\big)^2\big)\partial_x\big]u( x,\tau ), \label{FPK-NL-4}
\end{gather}
where coef\/f\/icients $\beta(\tau)$, $f(\tau)$, and $g(\tau)$ in
general case can depend on $X_u(\tau)$.

With the obvious notation $\Delta\xi=x$, equation (\ref{FPK-NL-4})
takes the form (\ref{FPK-NL-2}), (\ref{D-2}) when
$\beta(\tau)=X_u(\tau)$, and $f(\tau)$ and $g(\tau)$ do not depend
on  $X_u(\tau)$, and, therefore, (\ref{FPK-NL-4}) generalizes
(\ref{FPK-NL-2}), (\ref{D-2}).

From  (\ref{NORM-1}), (\ref{drift}), and (\ref{FPK-NL-4}) by
analogy with (\ref{EE-1}) we immediately obtain
\begin{gather}
\label{EE-A} \dot{X}_u(\tau)=-(\alpha -2a^2\epsilon)X_u(\tau)
-\varkappa \beta(\tau)
 + 2a\epsilon g(\tau).
\end{gather}
Similarly to equation (\ref{FPK-NL-3}), the Cauchy problem
(\ref{CAUCHY-1}) for the NFPE (\ref{FPK-NL-4}) induces the Cauchy
problem (\ref{EE-A}) and
\begin{gather}
\label{EE-B} X_u(s)=X_\gamma
\end{gather}
determining $X_u(\tau)$ independently of the solution of the FPE
(\ref{FPK-NL-4}), where $X_\gamma$ has the form~(\ref{EE-3}). When
$g(\tau)$ does not depend on  $X_u(\tau)$ and
$\beta(\tau)=X_u(\tau)$, we have the following explicit expression
for $X_u(\tau)$ from (\ref{EE-A}) and (\ref{EE-B})
\begin{gather*}
 X_u(\tau)=e^{-(\alpha+\varkappa -2a^2\epsilon)(\tau -s)} X_\gamma +  2a\epsilon\int_{s}^{\tau}e^{-(\alpha+\varkappa
-2a^2\epsilon)(\tau-\xi)} g(\xi)d\xi . 
\end{gather*}
The solution of equation (\ref{EE-A}) can be found also for a
different form of dependence of $g(\tau)$ on~$X_u(\tau)$. In any
case, f\/inding of the moment $X_u(\tau)$ separately from
solution of the nonlinear FPE   and substitution $X_u(\tau)$ into
equation (\ref{FPK-NL-4}) actually linearizes this equation with
respect to the PDF~$u(x,\tau)$.

In order to derive exact and approximate solutions it is helpful
to introduce  new variab\-les~$(y,\tau)$,
\begin{gather}
\label{variab-y} y= \frac{1}{a}\log\left(a x+
g(\tau)+\sqrt{f(\tau)+(a x+g(\tau))^2}\right),
\end{gather}
such that in the new variables the dif\/fusion term becomes a
constant. In the new variables equation (\ref{FPK-NL-4}) takes the
form
\begin{gather}
\label{FPK-NL-5}
\partial_\tau u( y,\tau )=\left[\alpha+ \displaystyle\frac{q(y,\tau)}{A_{(+)}(y,\tau)}\partial_y +\epsilon\partial_{yy}\right]
u(y,\tau).
\end{gather}
Here the following notations are used:
\begin{gather}
\label{NOTATION-q} q(y,\tau,\epsilon)=2\left(\varkappa \beta(\tau)
-
\displaystyle\frac{\alpha}{a}g(\tau)-\displaystyle\frac{\dot{g}(\tau)}{a}
\right) -\displaystyle\frac{\dot{f}(\tau)}{a}e^{-ay}+
\left(\displaystyle\frac{\alpha}{a} + a\epsilon \right)\,
A_{(-)}(y,\tau),
\\
\label{NOTATION-A} A_{(\pm)}(y,\tau)=e^{ay}\pm f(\tau)e^{-ay}.
\end{gather}

Let us impose special constraints on the functions $\beta(\tau)$,
$g(\tau)$, and $f(\tau)$ to reduce (\ref{FPK-NL-5}) to a~solvable
form and to construct exact solutions.

\subsection{Exact solution}

An exact solution of equation (\ref{FPK-NL-4}) can be found if the
functions $f(\tau)$ and $g(\tau)$  obey the conditions
\begin{gather}
\label{FPK-NL-6}   \dot f(\tau)+2 a c f(\tau)=0,\qquad  \dot
g(\tau)+\alpha  g(\tau) - a\varkappa\beta(\tau)=0, \qquad c=
\frac{\alpha}{a} + a\epsilon.
\end{gather}
In this case  $f(\tau)=f(s) e^{-2 a c (\tau -s)}$,
$g(\tau)=g(s)e^{-\alpha(\tau-s)}+a\varkappa\int_{s}^{\tau}e^{-\alpha(\tau-\xi)}\beta(\xi)d\xi$,
and equation~(\ref{FPK-NL-5}) takes the form
\begin{gather}
\label{FPK-NL-7}   \left[ -\partial_\tau+c\partial_y+\alpha
+\epsilon\partial_{yy}\right]u(y,\tau)=0.
\end{gather}
By changing the variables $(y,\tau)\rightarrow (z,\tau)$, where
$z=y+ c \tau$ and taking $v(z,\tau)=u(y,\tau)e^{-\alpha\tau}$, we
reduce (\ref{FPK-NL-7}) to the standard dif\/fusion equation
\begin{gather}
\label{FPK-NL-STANDARD} \left[
-\partial_{\tau}+\epsilon\partial_{zz}\right]v(z,\tau)=0,
\end{gather}
whose  Green function is well known
\begin{gather}
\label{FPK-NL-GREEN-STANDARD} G_{st}(\tau, s,z,z')=
\frac{1}{\sqrt{4\pi\epsilon(\tau-s)}}e^{-
\frac{(z-z')^2}{4\epsilon(\tau-s)}},
\end{gather}
where a f\/ixed value $\tau=s$ is taken as the initial time.

Then, for equation (\ref{FPK-NL-7}) we have
\begin{gather*}
G(\tau, s,y,y')=
\frac{e^{\alpha(\tau-s)}}{\sqrt{4\pi\epsilon(\tau-s)}}
\exp\left\{- \frac{1}{4\epsilon(\tau-s)}\big[y-y'+ c(\tau-s)
\big]^2\right\},
\end{gather*}
and for equation (\ref{FPK-NL-4})
\begin{gather}
  G(\tau, s,x,x')= \frac{e^{\alpha(\tau-s)}}{\sqrt{4\pi\epsilon(\tau-s)}}
\exp\left\{- \frac{1}{4\epsilon(\tau-s)}\Bigg[ \frac{1}{a}
\log \frac{ax+g(\tau)+\sqrt{f(\tau)+(ax+g(\tau))^2}}{ax'+g(s)+\sqrt{f(s)+(ax'+g(s))^2} } \right.\nonumber\\
\left.\phantom{G(\tau, s,x,x')=}{} +
c(\tau-s)\Bigg]^2\right\}\label{FPK-NL-GREEN-3}
\end{gather}
when conditions (\ref{FPK-NL-6}) are  satisf\/ied. Notice that the
Green function (\ref{FPK-NL-GREEN-3}) can be  considered as
kernel of the correspondent evolution operator of  the NFPE
(\ref{FPK-NL-4}), i.e.
\begin{gather*}\label{GREEN-NLIN-11}
u(x,\tau)=\widehat U (\tau, s, \gamma)(x)=
\int_{-\infty}^{+\infty}G(\tau, s,x,x')(f(\tau)+(a
x'+g(\tau))^2)^{-1/2} \gamma( x')d x'.
\end{gather*}
The evolution operator $\widehat U$ is nonlinear because the
functions $\beta(\tau)$, $f(\tau)$  and  $g(\tau)$ entering the
Green function (\ref{FPK-NL-GREEN-3}) depend on the function
$u(x',s)=\gamma(x')$ with the operator $\widehat U$ acting on it.

\subsection{Semiclassical approximation}

For equation (\ref{FPK-NL-5})   with the coef\/f\/icients
$\beta(\tau)$, $f(\tau)$,  and  $g(\tau)$ that are generally
dependent on~$\tau$ but independent of $\epsilon$ we construct a
leading term of  formal solution of the Cauchy problem and the
evolution operator asymptotic in small parameter $\epsilon$
($\epsilon\to 0$) in the class of trajectory-concentrated
functions.

Let us introduce  class $\mathcal
P_\epsilon^\tau(Y(\tau,\epsilon))$  of functions
\[
\Phi(y,\tau,\epsilon)=\varphi \left(\displaystyle\frac{\Delta
y}{\sqrt{\epsilon}},\tau,\epsilon \right),
\]
where $\Delta y=y-Y(\tau,\epsilon)$, $y$ is given by
(\ref{variab-y}), and $Y(\tau,\epsilon)$ is a functional parameter
of the class, regularly depends on $\epsilon$ and  is to be
determined. The function $\varphi (\xi,\tau,\epsilon )$ belongs to the Schwartz
space, smoothly
depends on $\xi$ and $\tau$,   decreases with $\xi\to\infty$, and
regularly depends on $\epsilon$ for $\epsilon\to 0$, i.e.\ it can
be expanded in the Taylor series around $\epsilon = 0$.

Functions of the class $\mathcal
P_\epsilon^\tau(Y(\tau,\epsilon))$ are normalized with respect to
the norm
\[
\|\Phi \|^2(\tau,\epsilon)= \int_{\mathbb R^1} {\Phi}^2
(\xi,\tau,\epsilon) d \xi .
\]
By analogy with \cite{BEL-TRIFONOV, BTS1} let us  call solutions
of the class $\mathcal P_\epsilon^\tau(Y(\tau,\epsilon))$ the
trajectory concentrated solutions. Restrict ourself with the case
when $Y(\tau,\epsilon)$ does not depend on $\epsilon$,
$Y(\tau,\epsilon)$= $Y(\tau)$. We  are interested here in
solutions of equation~(\ref{FPK-NL-5}) localized in $y$ in a
neighborhood of the curve $y=Y(\tau)$ for each f\/ixed
$\epsilon\in[0, 1[$ and $\tau\in\mathbb R^1$.

For the functions belonging to $\mathcal
P_\epsilon^\tau(Y(\tau))$, the following asymptotic estimates can
be imme\-dia\-tely obtained
\begin{gather*}
 \frac{\|(\Delta y)^k\Phi \|(\tau,\epsilon)}{\|\Phi \|(\tau,\epsilon)}=O(\epsilon^{k/2}), \qquad
 \frac{\|(\epsilon\partial_y)^k\Phi \|(\tau,\epsilon)}{\|\Phi \|(\tau,\epsilon)}=O(\epsilon^{k/2}),\\
 \frac{\|\epsilon(\partial_\tau +\dot Y(\tau,\epsilon)\partial_y)\Phi \|(\tau,\epsilon)}{\|\Phi \|(\tau,\epsilon)}=O(\epsilon).
\end{gather*}
Notice that $\xi= \frac{y-Y(\tau)}{\sqrt{\epsilon}}$ has the
estimate $\xi\sim O(1)$.

Transformation from $(y,\tau )$ to the variables $(\xi, \tau)$
reduces  equation (\ref{FPK-NL-5}) to the  form
\begin{gather}\label{FPK-NL-7a}
\widehat{L}\varphi(\xi,\tau,\epsilon)= \left[ -\partial_\tau
+\alpha +  \left(\dot{Y}(\tau) +
 \frac{q(\sqrt{\epsilon}\, \xi +Y(\tau),\tau)}{A_{(+)}(\sqrt{\epsilon}\,\xi +Y(\tau),\tau) }\right)\!
 \frac{1}{\sqrt{\epsilon}}\partial_{\xi}  +\partial_{\xi\xi}
\right]\varphi(\xi,\tau,\epsilon)=0,
\end{gather}
where notations (\ref{NOTATION-q}), (\ref{NOTATION-A}) are used.

Let us expand  coef\/f\/icients of equation (\ref{FPK-NL-7a}) and
the function $\varphi(\xi,\tau,\epsilon)$ into power series
in~$\sqrt{\epsilon}$, e.g.
\begin{gather}
\label{expan-1}
\varphi(\xi,\tau,\epsilon)=\varphi^{(0)}(\xi,\tau)+\sqrt{\epsilon}
\varphi^{(1)}(\xi,\tau)+\cdots .
\end{gather}
Equating coef\/f\/icients at equal powers of
$(\sqrt{\epsilon})^{k/2}$, $k=0,1,2,\dots$, in (\ref{FPK-NL-7})
we obtain the following equations determining  asymptotic solution
of equation (\ref{FPK-NL-7a})
\begin{gather}
  \dot Y(\tau)+ \frac{q_{0}\big(Y(\tau),\tau\big)}{A_{(+)}\big(Y(\tau),\tau\big)}=0, \label{asympt-1}\\
 \widehat L_{(0)}\varphi^{(0)}(\xi,\tau)=0, \label{asympt-2}\\
 \widehat
L_{(0)}\varphi^{(1)}(\xi,\tau)=- \widehat
L_{(1)}\varphi^{(0)}(\xi,\tau).  \label{asympt-3}
\end{gather}

Here we use the notations
\begin{gather*}
 \hat L_{(0)}=-\partial_\tau +\partial_{\xi\xi}+\xi R(\tau)\partial_\xi+\alpha, \\ 
 \hat L_{(1)}=\left(Q_0(\tau)\xi^2+Q_1(\tau)\right)\partial_\xi , \\ 
 R(\tau)=\displaystyle\frac{1}{A_{(+)}\big(Y(\tau),\tau\big)}\left[ \dot f(\tau)e^{-aY(\tau)}+\alpha A_{(+)}\big(Y(\tau),\tau\big)+
a \dot Y (\tau) A_{(-)}\big(Y(\tau),\tau\big)\right], \\ 
 Q_0(\tau)=  \frac{a}{2A_{(+)}\big(Y(\tau),\tau\big)}
\left[-\dot f(\tau)  e^{-aY(\tau)} + \alpha
A_{(-)}\big(Y(\tau),\tau\big) + a\dot
Y(\tau)A_{(+)}\big(Y(\tau),\tau\big)
\right]\nonumber \\
\phantom{Q_0(\tau)=}{} -  \frac{a
A_{(-)}\big(Y(\tau),\tau\big)}{A_{(+)}\big(Y(\tau),\tau\big)}R(\tau),
\qquad Q_1(\tau)= \frac{a
A_{(-)}\big(Y(\tau),\tau\big)}{A_{(+)}\big(Y(\tau),\tau\big)},
\\ 
  q_0(y,\tau)=q(y,\tau,\epsilon)|_{\epsilon=0} =
 2\left(\varkappa \beta(\tau) -  \frac{\alpha}{a}g(\tau)- \frac{\dot{g}(\tau)}{a} \right)
- \frac{\dot{f}(\tau)}{a}e^{-ay}+  \frac{\alpha}{a}  A_{(-)}(y,\tau). 
\end{gather*}
It is obvious that the construction of the  asymptotic solution is
reduced to the solution of equation (\ref{asympt-2}).

By changing dependent and independent variables in
(\ref{asympt-2}) according to the formulas
\begin{gather}
 \varphi^{(0)}(\xi,\tau)=e^{\alpha (\tau-s)} v(z,\tau'),\qquad \tau'=\int_{s}^{\tau}h^2(\eta)d\eta, \nonumber\\
z=h(\tau)\xi,\qquad h(\tau)=h(s)\exp\left(\int_s^\tau R(\eta)d\eta
\right),\label{changing-var-2}
\end{gather}
we reduce equation (\ref{asympt-2}) to equation
(\ref{FPK-NL-STANDARD}) with $\epsilon =1$ whose Green function is
given  by (\ref{FPK-NL-GREEN-STANDARD}), i.e.
\begin{gather}
\label{GREEN-11} G_{st}(\tau', 0,z,z')=
\frac{1}{\sqrt{4\pi\tau'}}e^{- \frac{(z-z')^2}{4\tau'}}.
\end{gather}
Notice that in (\ref{changing-var-2}) $\tau=s$ corresponds to
$\tau'=0$. Therefore, the solution of the Cauchy problem
$\big(-\partial_{\tau'}+\partial_{zz}\big)v(z,\tau')=0$, $
v(z,0)=v_0(z)$ given by  the Green function (\ref{GREEN-11}) reads
\begin{gather}
\label{CAUCHY-II-1}
v(z,\tau')=\int_{-\infty}^{\infty}G_{st}(\tau',0,z,z')v_0(z')dz'
\end{gather}
and for the problem  $\big(-\partial_{\tau'}+\partial_{zz}\big)
u(z,\tau')=F(z,\tau')$, $u(z,0)=0$, we have
\begin{gather}
\label{DUHAMEL-II-2}
u(z,\tau')=\int_{0}^{\tau'}d\eta\int_{-\infty}^{\infty}
G_{st}(\tau'-\eta,0,z,z')F(z',\eta)dz'.
\end{gather}

Let us now f\/ind the asymptotic solution of the Cauchy problem
\begin{gather}
\label{CAUCHY} \varphi(\xi,s)=\gamma(\xi)
\end{gather}
for equation (\ref{FPK-NL-7a}) in the form (\ref{expan-1}). To
this end we change variables from $(z,\tau')$ to $(\xi,\tau)$
in~(\ref{CAUCHY-II-1}) according to (\ref{changing-var-2}) and
obtain the solution of the Cauchy problem
$\varphi^{(0)}(\xi,s)=\gamma(\xi)$ for equation~(\ref{asympt-2})
\begin{gather}
 \varphi^{(0)}(\xi,\tau)=\int_{-\infty}^{\infty}e^{\alpha(\tau-s)}
G_{st}(\tau'(\tau),0,h(\tau)\xi,h(s)\xi')\gamma(\xi')h(s)d\xi' \nonumber\\
\phantom{\varphi^{(0)}(\xi,\tau)}{}  =\int_{-\infty}^{\infty}
\frac{e^{\alpha(\tau-s)}}{\sqrt{4\pi \int_s^\tau h^2(\eta)d\eta}}
\exp\left[- \frac{\big( h(\tau)\xi-h(s)\xi' \big)^2}{4 \int_s^\tau
h^2(\eta)d\eta} \right]\gamma(\xi')h(s)d\xi'.\label{asympt-10}
\end{gather}
Here the relations $z'=h(s)\xi'$,
$\gamma(\xi')=v_0(z')=v_0(h(s)\xi')$ are  used. Similarly, in
accordance with (\ref{DUHAMEL-II-2}) for the function
$\varphi^{(1)}(\xi,\tau)$ determined by (\ref{asympt-3}) we have
\begin{gather}
\label{DUHAMEL-II-3}
\varphi^{(1)}(\xi,\tau)=\int_{s}^{\tau}d\eta\int_{-\infty}^{\infty}
e^{\alpha(\tau-\eta)}G_{st}(\tau'(\tau)-\tau'(\eta),0,h(\tau)\xi,h(\eta)\xi')\tilde
F(\xi',\eta) h(\eta)d\xi',
\end{gather}
where $F(h(\eta)\xi',\tau'(\eta))=e^{-\alpha(\eta
-s)}h^{-2}(\eta)\tilde F(\xi',\eta)$, $\tilde F(\xi,\tau)=-\hat
L_{(1)}\varphi^{(0)}(\xi,\tau)$,
\begin{gather}
\label{DUHAMEL-II-4}
\varphi^{(1)}(\xi,\tau)=\int_{s}^{\tau}d\eta\int_{-\infty}^{\infty}
\displaystyle\frac{e^{\alpha(\tau-\eta)}}{\sqrt{4\pi
\int_\eta^\tau h^2(y)dy}} \exp\left[- \frac{\big(
h(\tau)\xi-h(\eta)\xi' \big)^2}{4 \int_\eta^\tau h^2(y)dy} \right]
\tilde F(\xi',\eta) h(\eta)d\xi'.
\end{gather}

Thus, the function (\ref{expan-1}) determined by expressions
(\ref{asympt-1}), (\ref{asympt-10}), (\ref{DUHAMEL-II-3}) gives
the asymptotic solution of the Cauchy problem (\ref{FPK-NL-7a}),
(\ref{CAUCHY})
 accurate to $O(\epsilon)$
in the following meaning. Denote the action of the operator
$\widehat L$ def\/ined by (\ref{FPK-NL-7a}) on the function
$\varphi^{(0)}(\xi , \tau)+\sqrt{\epsilon} \varphi^{(1)}(\xi ,
\tau)$ as
\begin{gather}
\label{approx-1} g^{(2)}(\xi,\tau,\epsilon):=\widehat
L\big(\varphi^{(0)}(\xi , \tau)+\sqrt{\epsilon} \varphi^{(1)}(\xi
, \tau)\big),
\end{gather}
then
\begin{gather}
\label{approx-2}  \frac{\|g^{(2)} \|(\tau,\epsilon)
}{\|\varphi\|(\tau,\epsilon)}=O(\epsilon).
\end{gather}
Similarly to (\ref{GREEN-NLIN}), we can consider  the solution of
the Cauchy problem as action of the evolution operator  of the
operator $\widehat L$  on the function $\gamma(\xi)$.

\section{Concluding remarks}

Asset interactions at the market are naturally supposed to be
complex and nonlinear. The  approach discussed in the work is
intended for the investigation of the asset market by using
simple nonlinear model directly generalizing known models based on
the linear FPE (\ref{FPK-FRIED}). The asymptotics obtained,
equations (\ref{expan-1}), (\ref{asympt-10}),
(\ref{DUHAMEL-II-4}), can be considered as a necessary step in
studying of nonlinear properties of the asset market.

The approach provides the way to analyze ef\/fects of nonlinearity
on the distribution dynamics of asset price increments at the
stock market or  currency exchange. Such an analysis necessarily
implies the statistical processing of corresponding empirical
data to estimate values of the model parameters, which is beyond
the scope of the present work and is a subject of a special
research.

Following \cite{FRIEDRICH, SORNETTE, SORNETTE2}, we use quadratic
approximation for  the dif\/fusion coef\/f\/icient (\ref{D-2}) and
li\-near drift  (\ref{D-1}).  Our formalism  can be also applied
to the NFPE with more general dependence of coef\/f\/icients on
$\Delta \xi$. The quadratic approximation of the dif\/fusion
coef\/f\/icient (\ref{D-2}) is  a  generalization of the constant
dif\/fusion that results in non-Gaussian probability density
functions. These PDFs are  asymptotic solutions of the NFPE
(\ref{FPK-NL-4}) in semiclassical approximation  in explicit form
given by (\ref{expan-1}), (\ref{asympt-10}), (\ref{DUHAMEL-II-4})
accurate to $O(\epsilon)$ in accordance with (\ref{approx-1}),
(\ref{approx-2}). The construction of semiclassical solutions is
based on exact solution of the NFPE with constant dif\/fusion
(\ref{FPK-NL-3}).

Preliminary calculations  show that for the initial function
$\gamma(x)$ having the power law asymptotics (``fat tails'') the
distribution keeps the non-Gaussian shape during its evolution.

In conclusion we would notice that since the  stochastic
properties of asset price increments are studied ef\/fectively by
numerical simulation, it is of interest to formulate stochastic
dif\/ferential equations with  the nonlinear feedback related to
the NFPE considered above.

\subsection*{Acknowledgements}
The work was supported by  the President of the Russian
Federation, Grant  No SS-871.2008.2 and  Grant of the RFBR No
07-01-08035. A.~Shapovalov acknowledges the partial support from
the International Mathematical Union for participation in the
Seventh International Conference ``Symmetry in Nonlinear
Mathematical Physics''.

\pdfbookmark[1]{References}{ref}
\LastPageEnding

\end{document}